# Comparative Analysis of Large Load Interconnection Requirements Across U.S. Transmission Entities

**R. ZAHEDI, R. ANILKUMAR, A. ZAMANI**

**Danovo Energy Solutions LLC**
**USA**

## SUMMARY

The rapid growth of large loads, especially data centers driven by artificial intelligence, is creating unprecedented demands on the North American electric grid. Data centers now account for nearly 80% of large load interconnection requests in the Western Interconnection, and U.S. electricity consumption from data centers is projected to reach 325–580 TWh by 2028. Despite this urgency, technical interconnection requirements for large loads have not yet matured and still contain several gaps. This paper presents a comparative analysis of large load interconnection requirements across eleven U.S. entities, including AEP, Ameren, CAISO, Dominion, Duke Energy, ERCOT, LIPA, MISO, PJM, Southern Company, and SPP, under seven technical categories: large load definition, load data and modeling, fault ride-through, load curtailment and shedding, dispatchability, power quality, and reactive power control. The analysis reveals significant variation in both the scope and specificity of requirements. LIPA addresses all seven technical categories at least partially, while Dominion and Duke Energy provide detailed coverage across nearly all categories. The analysis also finds that data center applicants can be viewed as potential grid resources, rather than only as consumers. The findings provide a structured baseline from which utilities, regulators, ISOs/RTOs, and developers can identify gaps, benchmark against peers, and accelerate the development of interconnection standards.

RZahedi@danovoenergy.com

## 1. INTRODUCTION

The North American electric grid is undergoing a rapid transformation driven by the unprecedented growth of large electrical loads including data centers, cryptocurrency mining facilities, and heavy-duty electric vehicle charging stations [1], [2]. Among these, data centers have emerged as the dominant source of large load growth, fueled by the explosive expansion of artificial intelligence [3]. Data center load interconnection requests comprise nearly 80% of the large load interconnection across the Western Interconnection [4]. U.S. Department of Energy report shows data centers consumed about 4.4% of total U.S. electricity in 2023 [5]. It is expected that data centers consume between 6.7%-12% of the of total U.S. electricity [6]. In 2024, global data generation exceeded 149 zettabytes, and U.S. electricity demand from data centers rose from 58 TWh in 2014 to an estimated 176 TWh in 2023, with the U.S. Department of Energy projecting consumption of 325–580 TWh by 2028 [1], [3].

The scale and speed of these interconnection requests overwhelm utility processes and expose reliability gaps. A July 2024 incident in the Eastern Interconnection illustrated these risks as a 230 kV transmission line fault triggered loss of approximately 1,500 MW of voltage-sensitive data center load [7]. During the event, the large load was transferred to backup generation and caused frequency overshoot and high voltages. Similar events involving cryptocurrency mining loads in ERCOT and the Eastern Interconnection, where 25 individual load losses of 100–400 MW each were recorded between late 2023 and early 2025, which further emphasize the importance of the problem [1].

The integration of large electrical loads, such as data centers, into the power system introduces significant technical and operational challenges. As described in a recent IEEE report, grid interconnections is a key issues associated with data centers and other large loads [8].

The regulatory landscape is also evolving rapidly. In October 2025, the U.S. Department of Energy directed the Federal Energy Regulatory Commission to initiate rulemaking to standardize and accelerate large load interconnections to the interstate transmission system. Concurrently, regional operators such as CAISO, ERCOT, SPP, and PJM are developing their own proposals. Despite this activity, current interconnection requirements for large loads remain fragmented across U.S. transmission entities. Definitions of a "large load," voltage ride-through (VRT) and frequency ride-through (FRT) expectations, load curtailment provisions, dispatchability, power quality, and reactive power requirements vary significantly from one entity to another [1].

This paper addresses that gap by providing a comparative analysis of large load interconnection requirements across eleven U.S. transmission entities with a focus on technical performance requirements at the point of interconnection (POI). This analysis could extract key findings and numerical values that can be used to fill the knowledge gaps. The analysis draws on publicly available utility handbooks, ISO/RTO guidance documents, and proposed standards to identify areas of convergence and divergence, and highlight gaps in current practice. The methodology combines cross-entity comparisons and provides an assessment of requirements maturity among entities to provide a practical reference for utilities, regulators, and developers.

## 2. COMPARATIVE FINDINGS OF U.S. TRANSMISSION ENTITIES

This section presents a comparative analysis of large load interconnection requirements across eleven U.S. transmission entities: AEP [9], Ameren [10], [11], CAISO [12], [13], Dominion [14], Duke Energy [15], [16], ERCOT [17], [18], LIPA [19], [20], MISO [21], [22], PJM [23], [24], Southern Company [25], [26], [27], and SPP [28], [29], [30]. The analysis is organized into seven technical categories following the paper's structure: large load definition; load data, modeling and study process; fault ride-through (VRT and FRT); load curtailment and load shedding; dispatchability; power quality; and reactive power control. For each category, entity-specific requirements are compared side by side to identify areas of convergence and divergence, as well as take out key findings.

### 2.1 Large load definition

Almost all entities agree that policies are needed for the interconnection of large loads. However, as shown in Fig. 1, there is no consistent definition of a large load across entities. PJM, MISO, ERCOT, AEP, Dominion, and LIPA apply large-load requirements to facilities whose demand exceeds a specified power threshold. By contrast, SPP, Southern Company, and Duke Energy use two criteria: load size and interconnection voltage.

The lowest threshold is 10 MW, which applies only under SPP's requirements for loads connected at 69 kV. Facilities below this level fall outside every surveyed entity's large-load classification. LIPA establishes the next-lowest threshold at 20 MW, while 50 MW is the most commonly adopted threshold among the surveyed entities. The highest threshold is 75 MW. Therefore, any facility with a demand of 75 MW or greater would be classified as a large load by all surveyed entities**.**

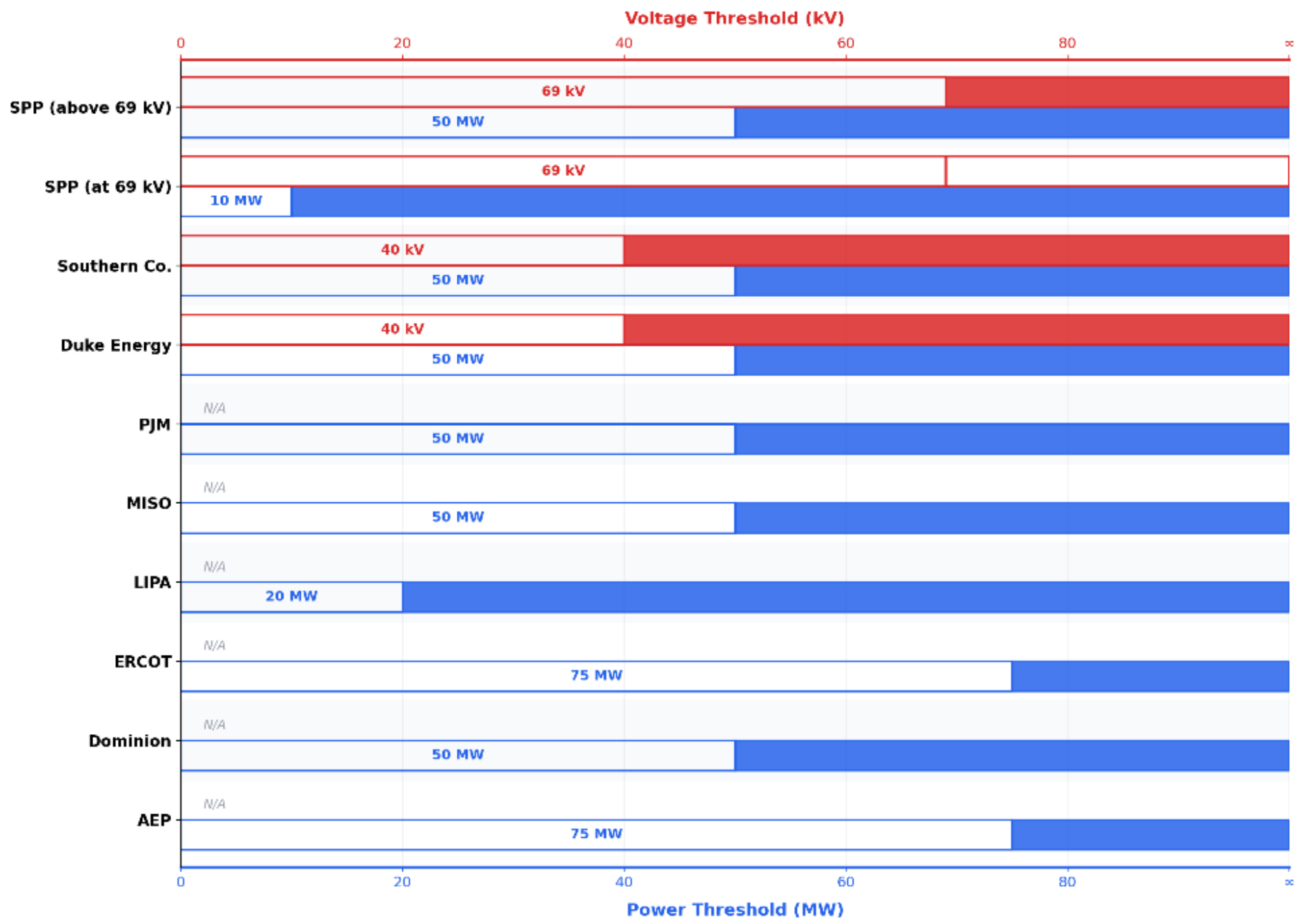

**Fig. 1. Large Load MW Thresholds Across Entities**

### 2.2 Load data and modeling

Without accurate composition data, planners cannot predict the large load facility behavior. Equally important are the simulation tools used to assess these impacts, and the study processes that define scope, timeline, and acceptance criteria for energization.

*Load composition data*

NERC issued a Level 3 Essential Action Alert for computational loads interacting with the power grid on May 4$^{th}$, 2026 [31]. The alert asks nine pieces of modeling data that entities are expected to collect from large load facility developers. Table I shows the items these entities are asking as of now based on their published documents.

**Table I. Load Composition Requirements Comparison**

| Modeling data | AEP | Ameren | CAISO | Dominion | Duke Energy | ERCOT | LIPA | MISO | PJM | Southern Co. | SPP |
|---|---|---|---|---|---|---|---|---|---|---|---|
| **Load Size on a Seasonal Basis** | ✓ | ✓ | | ✓ | ✓ | | ✓ | ✓ | ✓ | ✓ | |
| **Power Factor on a Seasonal Basis** | ✓ | ✓ | | ✓ | ✓ | | ✓ | ✓ | | ✓ | |
| **Buildout Schedule** | ✓ | ✓ | | ✓ | ✓ | ✓ | | ✓ | ✓ | ✓ | |
| **Dynamic Characteristics** | ✓ | | ✓ | ✓ | ✓ | ✓ | ✓ | ✓ | | ✓ | |
| **Load Composition (IT vs non-IT / PUE)** | ✓ | | | ✓ | ✓ | | ✓ | ✓ | | ✓ | |
| **Ramp Rate** | | | ✓ | ✓ | ✓ | | ✓ | ✓ | | ✓ | ✓ |
| **Protective Devices** | ✓ | ✓ | | ✓ | ✓ | | ✓ | | | ✓ | |
| **On-Site Generation** | | | ✓ | ✓ | ✓ | ✓ | ✓ | ✓ | ✓ | ✓ | ✓ |
| **Facility Use** | ✓ | | | | | | | | | ✓ | |

*Load model requirements*

Table II compares the study types and processes across entities. All entities require steady-state and short-circuit studies as a baseline, with dynamic/transient stability and EMT studies added as system conditions warrant.

**Table II. Load Data, Modeling, and Study Process Requirements Comparison**

| **Entity** | **Study Types** | **Model Format** |
|---|---|---|
| AEP | Steady-state, short-circuit, dynamic/transient stability, facility study | PSS/E accepted; PSCAD may be requested |
| Ameren | Power flow; adds short-circuit, stability, switching, or EMT as needed | NA |
| CAISO | Steady-state, dynamic, short-circuit, EMT (if requested) | NA |
| Dominion | Steady-state, short-circuit, dynamic stability, EMT (if requested) | PSS/E required; PSCAD may be requested |
| Duke Energy | Steady-state, short-circuit, dynamic, EMT, harmonic | NA |
| ERCOT | Steady-state, stability, short-circuit, Subsynchronous Oscillation (SSO) | Dynamic model per TSP/ERCOT format |
| LIPA | Power flow/voltage drop, short-circuit, stability | NA |
| MISO | Steady-state, stability, short-circuit, EMT (if triggered), SSO screening | PERC model (positive-seq); EMT model |
| PJM | Included in RTEP reliability studies | NA |
| Southern Co. | Steady-state, stability, short-circuit, EMT (if requested) | NA |

## 2.3 Fault ride-through

Fault ride-through refers to the combined capability and obligation of a facility to remain connected at the POI through voltage and frequency disturbances. The following subsections address VRT and FRT separately.

*Voltage ride-through*

Among the eleven entities, five (ERCOT, SPP, LIPA, MISO, and CAISO) publish explicit numeric VRT envelopes with time–voltage bands. AEP references the ITIC/CBEMA curve, a widely used equipment voltage-tolerance standard, and requires ride-through of at least two consecutive voltage excursions outside 0.92–1.05 pu.

MISO's proposed VRT framework, shown in Fig. 2, is aligned with IEEE 2800 and NERC PRC-029. Also, MISO uniquely defines active power boundaries by voltage region: in the continuous range (0.90–1.20 pu), no power reduction is permitted; in the mandatory recovery range (0.50–0.90 pu), proportional active power reduction is allowed; and in the stability priority range (below 0.50 pu), flexible power reduction is allowed to maintain synchronization. The facility must also withstand three voltage disturbances within 10 seconds and ride through sudden phase-angle changes up to 25 degrees.

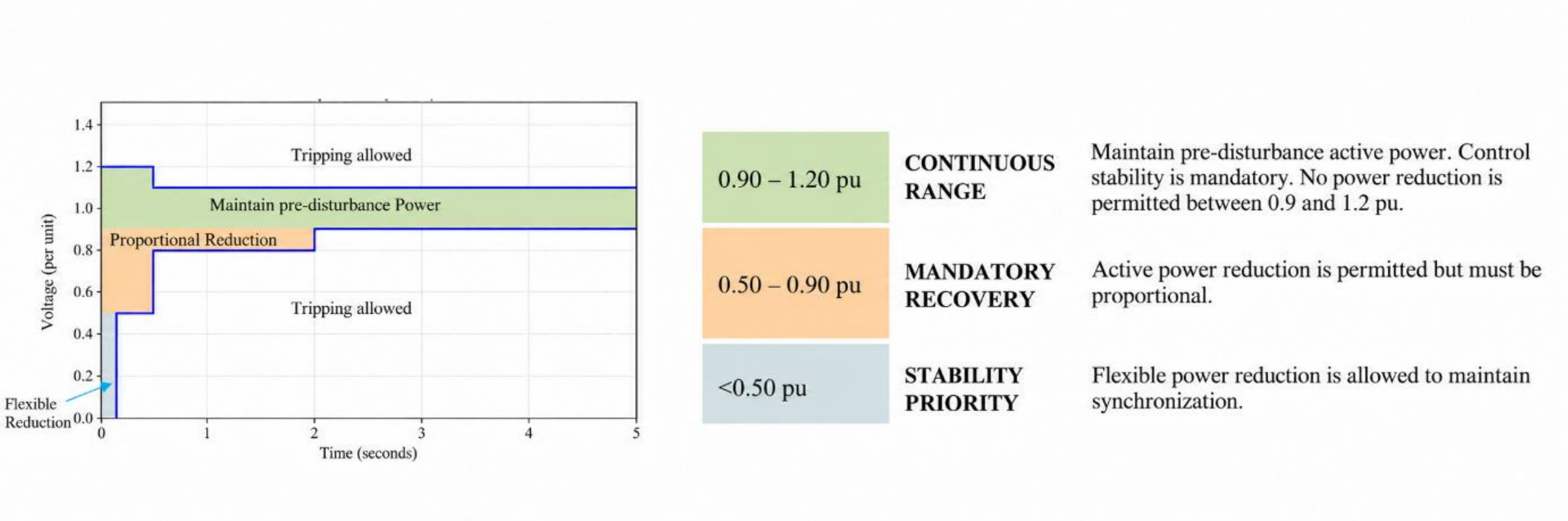

**Fig. 2. MISO proposed VRT envelope with voltage boundaries (left) and active power boundaries (right).**

ERCOT's proposed VRT for large loads requires continuous operation at 0.90–1.10 pu, with staged ride-through outside that range. During voltage disturbances, current consumption must not exceed 150% of the maximum during normal operations. That extra current is allowed for no more than 0.5 seconds. For $0.80 \leq V < 0.90$, large load shall continue consuming active power during the disturbance. For $V < 0.80$, large load may temporarily reduce active-power consumption. Also, the large load shall return to at least 90% of its pre-disturbance consumption within 2 seconds after voltage returns above 0.90 p.u. SPP's VRT requirements are aligned with the ITIC/CBEMA curves and IEEE 2800-2022, and require the facility to remain connected for a minimum of six voltage fault-clearing attempts within a 90-second period. Constant-current control is required for large loads during disturbances; constant-power control is prohibited. Dominion takes a reclosing-tolerance approach, requiring ride-through of at least six automatic reclosing attempts of 50–70 ms each and recommending undervoltage pickup settings of approximately 80–120 ms or more. Duke Energy and Southern Company require the facility to submit VRT settings for review, with the utility retaining authority to mandate adjustments. LIPA specifies an explicit numeric VRT envelope, as listed in Table III. CAISO defines a comparable voltage ride-through envelope, evaluated at the high side of the main power transformer. CAISO requires post-fault active-power recovery, with loads restoring at least 90% of their pre-disturbance active power within three seconds after the point-of-interconnection voltage recovers to 0.90 pu or above.

Fig. 3 provides a side-by-side bar chart comparison of the five numeric VRT envelopes. All five entities expect a normal operating interval of 0.90–1.10 pu and permit tripping above 1.20 pu. Beyond that band, however, the entities diverge significantly in how much time they expect loads to ride through during disturbances. In voltage sag conditions, CAISO requires the longest ride-through times; CAISO requires up to 6.0 s in the 0.70–0.90 pu band, roughly 3× the 2.0 s required by ERCOT, SPP, MISO, and LIPA, reflecting an emphasis on staying connected through deep voltage dips. Conversely, in voltage swell conditions, LIPA, and CAISO require 2.0 s in the 1.10–1.20 pu band, 2–4× more than MISO and SPP, indicating a stronger emphasis on overvoltage resilience.

**Table III. Voltage Ride-Through Requirements for ERCOT, SPP, LIPA, MISO, and CAISO**

| Voltage (pu) | ERCOT | SPP | LIPA | MISO | CAISO |
|---|---|---|---|---|---|
| V > 1.20 | May trip | May trip | May trip | May trip | May trip |
| 1.10 < V ≤ 1.20 | 1.0 s | 0.50 s | 2.0 s | 0.50 s | 2.0 s |
| 0.90 ≤ V ≤ 1.10 | Continuous | Continuous | Continuous | Continuous | Continuous |
| 0.80 ≤ V < 0.90 | 2.0 s | 2.0 s | 2.0 s | 2.0 s | 6.0 s |
| 0.70 ≤ V < 0.80 | 0.50 s | 0.50 s | 0.50 s | 0.50 s | 6.0 s |
| 0.50 ≤ V < 0.70 | 0.50 s | 0.50 s | 0.50 s | 0.50 s | 3.0 s |
| 0.35 ≤ V < 0.50 | 0.25 s | 0.15 s | 0.15 s | 0.15 s | 1.2 s |
| 0.25 ≤ V < 0.35 | 0.15 s | 0.15 s | 0.15 s | 0.15 s | 1.2 s |
| V < 0.25 | 0.15 s | 0.15 s | 0.15 s | 0.15 s | 0.15 s |

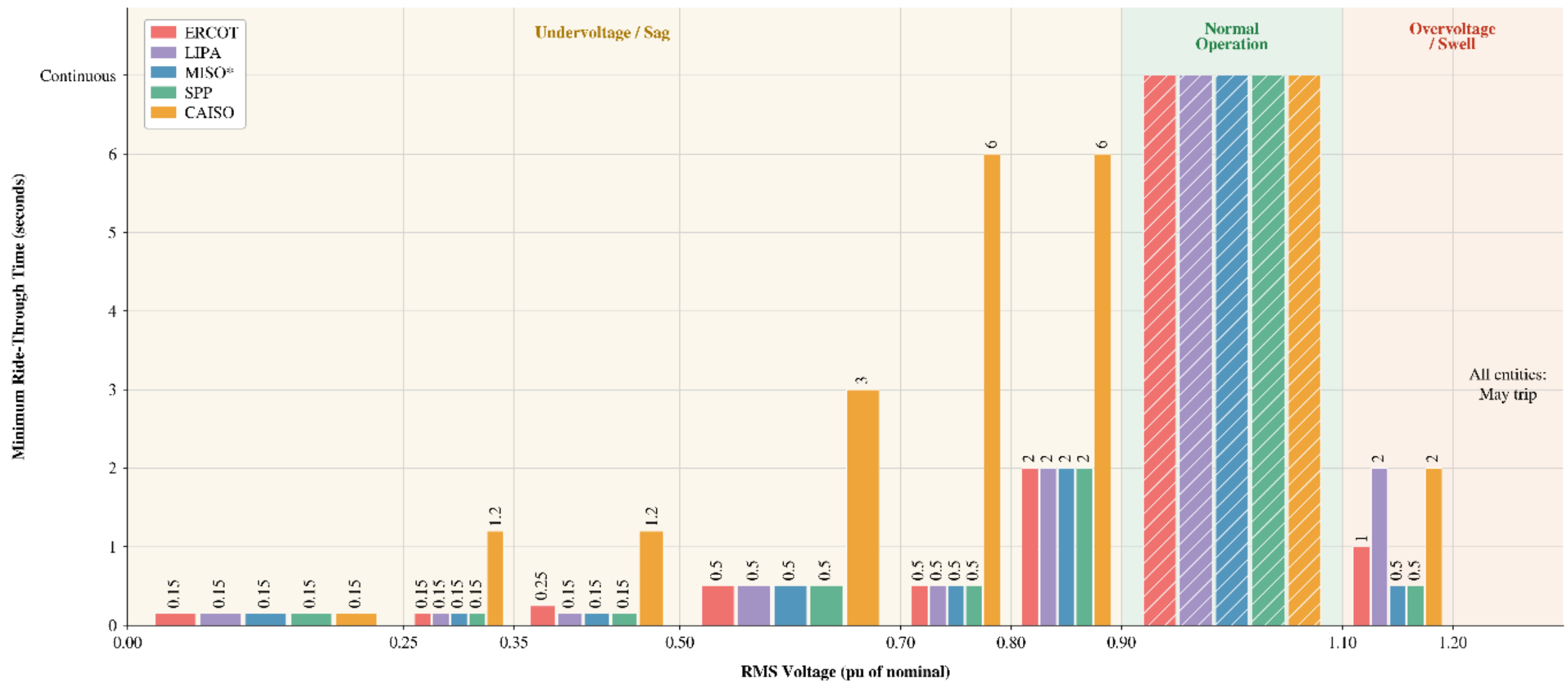

**Fig. 3. Voltage ride-through comparison: minimum ride-through time at each voltage level for ERCOT, SPP, LIPA, MISO, and CAISO.**

*Frequency ride-through*

FRT requires facilities to remain connected through defined frequency deviations and durations. For converter-dominated data centers, simultaneous tripping can remove hundreds of megawatts, motivating explicit FRT requirements beyond system-wide under-frequency load shedding schemes. Among the eleven entities, ERCOT, SPP, LIPA, MISO, and CAISO publish explicit numeric FRT envelopes. Except ERCOT, the remaining four are identical as shown in Table IV. ERCOT requires the largest FRT band, and continued grid-current consumption. Also, it states that active power should remain within 10% of the pre-disturbance level.

**Table IV. Frequency Ride-Through Requirements for ERCOT, SPP, LIPA, MISO, and CAISO**

| Frequency (Hz) | ERCOT | SPP / LIPA / CAISO/MISO |
|---|---|---|
| f > 63.0 | May trip | May trip |
| 61.8 < f ≤ 63.0 | 299 s | May trip |
| 61.2 < f ≤ 61.8 | 299 s | 299 s |
| 58.8 ≤ f ≤ 61.2 | Continuous | Continuous |
| 57.5 ≤ f < 58.8 | 299 s | 299 s |
| 57.0 ≤ f < 57.5 | May trip | 299 s |
| f < 57.0 | May trip | May trip |

MISO’s proposed FRT framework, shown in Fig. 4, including the active power boundaries that accompany the frequency bands. In the continuous range (58.8–61.2 Hz), the facility must maintain pre-disturbance active power. In the extended ride-through bands (57.0–58.8 Hz and 61.2–61.8 Hz), the facility must remain connected for 299 seconds and maintain active power within ±10% of pre-disturbance levels. MISO further requires ride-through when the rate of change of frequency (RoCoF) is within ±5.0 Hz/s.

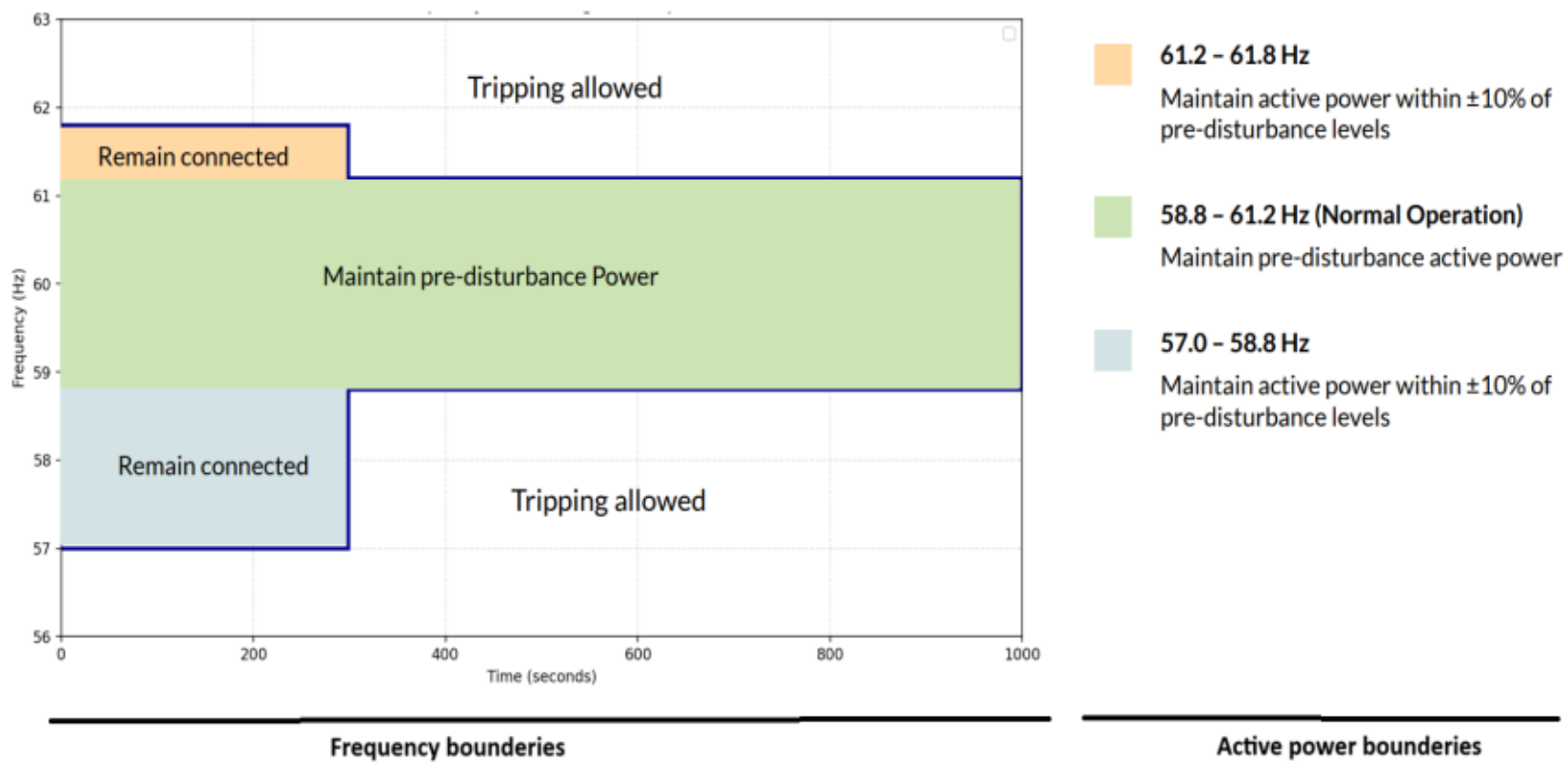

**Fig. 4. MISO proposed FRT envelope with frequency (left) and active power boundaries (right).**

## 2.4 Load curtailment and load shedding

Load curtailment and shedding provisions determine how transmission entities can ask large loads to reduce consumption during emergencies. Table V summarizes coverage across entities for three categories: under-frequency load shedding (UFLS), under-voltage load shedding (UVLS), and load curtailment.

**Table V. Load Curtailment and Load Shedding Requirements Summary**

| Entity | UFLS | UVLS | Curtailment | Key Details |
|---|---|---|---|---|
| AEP | Yes | — | — | UFLS 59.50–58.50 Hz in steps |
| Ameren | Yes | Yes | — | UFLS: 10% shed at 59.3, 59.0, 58.7 Hz |
| CAISO | Yes | Yes | Yes | Require 24/7 curtailment capability. Specific load-shedding is under development |
| Dominion | Yes | — | Yes | Requires data center applicants to disclose whether backup generation or BESS can participate in demand-side management |
| Duke Energy | Yes | — | Yes | UFLS required for ≥ 30% of peak capacity |
| ERCOT | Yes | — | — | — |
| LIPA | Yes | — | Yes | During emergency conditions, the facility operator shall reduce end-user load as directed by the LIPA System Operator<br>UFLS: if POI < 59.5 Hz, facility active-power demand must be reduced by at least the lesser of 17% of Maximum Rated Demand or 50% of pre-event active-power demand, within 0.30 s |
| MISO | — | — | — | Large Loads may initially be allowed without UFLS/UVLS, but MISO reserves the right to require installation at customer's expense. |
| PJM | — | — | Yes | Load curtailment during supply shortages |
| Southern Co. | Yes | Yes | Yes | May include in Manual Load Shed, UVLS, and/or UFLS at sole discretion |

Southern Company and CAISO are the only entities with requirements spanning all three categories, including the authority to include large loads in manual load shed, UVLS, and UFLS programs at its sole discretion. The details of the requirements for each entity are described in Table V. LIPA's updated requirements now add a quantitative UFLS obligation: when the POI frequency falls below 59.5 Hz, the facility must reduce active-power demand by at least the lesser of 17% of maximum rated demand or 50% of pre-event demand within 0.30 s. Also, as mentioned in the Table, Dominion requires data center applicants to disclose whether backup generation or battery energy storage system (BESS) can participate in demand-side management. This means data center facilities can be seen as potential grid resources, not just consumers.

## 2.5 Dispatchability

Dispatchability requirements govern how quickly and how variably a large load may change its demand, and how it must limit oscillatory behavior to preserve grid stability.

*Fluctuation and ramp-rate*

Southern Company, Duke Energy, MISO, SPP, LIPA, and CAISO have explicit ramp rate limits. As shown in Table VI, Southern Company provides the most detailed framework with an exact fluctuation constraint description, while MISO and LIPA have minor constraints on the fluctuation, and Duke Energy, SPP, and CAISO have not published a numerical fluctuation constraint.

**Table VI. Fluctuation and Ramp-Rate Constraints Comparison**

| Entity | Ramp Rate Limit | Intra-Minute Fluctuation Constraint |
|---|---|---|
| Southern Co. | 20 MW/min | • The median change in active power of every 4-6 seconds should not exceed the higher of 1 MW or 1.0% of the maximum capacity of the load.<br>• The top quartile (P75) change in active power every 4-6 seconds should not exceed 10 MW |
| Duke Energy | 10 MW/min | NA |
| MISO | 30 MW/min | 10 MW max change in a sliding 5-second window. |
| SPP | 20 MW/min | NA |
| LIPA | 20 MW/min | Measured as change over any 60-second period |
| CAISO | 20 MW/min | Measured as change over any 10-min period |

*Load oscillation*

Load oscillation constraints address the risk that cyclical or variable demand patterns from large loads, particularly data centers with power-electronic loads, may excite natural oscillatory modes of the interconnected power system. Four frequency bands are of concern: sub-synchronous (0.01–5 Hz), inter-area (0.1–0.5 Hz), local plant (1–5 Hz), and super-synchronous (5–55 Hz). Southern Company and LIPA publish explicit FFT-based active-power oscillation limits. The measurement method requires FFT analysis of three-phase active-power time series at the POI with 0.1 Hz bin resolution corresponding to a 10-second sampling window, with a Hann window and zero-padding recommended to reduce spectral leakage [25]. LIPA likewise defines explicit zero-to-peak oscillation limits, capping the sum of any two adjacent bins at 10 MW and the sum of all bins at 20 MW in the 0.1–5 Hz range, with a separate 3.5 MW cap in the 5–55 Hz range, all measured over rolling windows at the POI, as detailed in Table VII. Duke Energy references the inter-area frequency range (0.01–0.5 Hz) and requires that the facility be designed to minimize load oscillation in this band and must not create adverse or unstable interactions with other controls or the transmission network [15].

**Table VII. Load Oscillation Constraints Comparison**

| Entity | Quantitative Limits | Measurement Method |
|---|---|---|
| Southern Co. | 0.1 < f < 5 Hz: 10 MW max FFT magnitude at each 0.1 Hz bin.<br>5 ≤ f ≤ 55 Hz: 2.5 MW max FFT magnitude at each 0.1 Hz bin. | FFT of three-phase active power measured at the POI<br>0.1 Hz bin resolution using a 10-second sampling window<br>Hann window with zero-padding recommended |
| LIPA | 0.1 < f < 5.0 Hz: 10.0 MW zero-to-peak for the sum of any two adjacent 0.1 Hz bins (60 s rolling window)<br>0.1 < f < 5.0 Hz: 20.0 MW zero-to-peak for the sum of all bins (60 s rolling window).<br>5.0 ≤ f ≤ 55.0 Hz: 3.5 MW zero-to-peak for the sum of any two adjacent 1.0 Hz bins (10 s rolling window). | FFT of three-phase active power measured at the POI<br>0.1 < f < 5.0 Hz : sampled at ≥ 60 Hz, 10 s FFT window, 0.1 Hz bins, averaged over a 60 s rolling window<br>5.0 ≤ f ≤ 55.0 Hz : sampled at ≥ 600 Hz, 1 s FFT window, 1.0 Hz bins, averaged over a 10 s rolling window<br>Facility Owner shall disclose the electrical characteristics of any processes creating cyclic variability of active power demand within a 10-minute period that exceeds more than 5% of Maximum Rated Demand. |
| Duke Energy | Must minimize oscillation in 0.01–0.5 Hz range. Must not create adverse or unstable interactions with other controls or transmission network. | ≥ 60 samples/second, satellite-synchronized clock |

## 2.6 Power quality

Power quality requirements at the POI includes three principal categories: harmonic distortion, voltage flicker, and voltage unbalance. Coverage is uneven across entities, with harmonics being the most widely addressed and voltage unbalance the least.

*Harmonic distortion*

Six of eleven entities address harmonics distortion. IEEE 519 is the common standard across the board, but what each entity layers on top varies quite a bit, as detailed in Table VIII.

**Table VIII. Harmonic Distortion Requirements Comparison**

| Entity | Reference Std. | Additional Requirements / Notes |
|---|---|---|
| Dominion | IEEE 519 | Size cap: Inter-harmonic currents limited to 25% of IEEE 519 values.<br>THD or TDD must include inter-harmonics (10 Hz increments).<br>Current distortion must not exceed IEEE 519 limits more than 5% of the time (one-week probability).<br>Dedicated PQ meters at POI measuring harmonics through the 50th and THD for voltage and current. |
| AEP | IEEE 519-2022 | Evaluated at POM using a class-A PQ meter with per-phase voltage and current inputs. |
| Duke Energy | Entity-specific | Current injection limits by voltage class and $I_{sc}/I_{load}$ ratio (released as a chart)<br>Even harmonics = 25% of odd-harmonic limits.<br>DC offset not allowed.<br>Exceeding limits for 1 hr/day or 150% for > 5 min on any 2 days in 6 months is not allowed. |
| LIPA | IEEE 519 | Strong systems (SCR > 20): current limits — h<11: 2.0%, h≥11: 1.0%, TRD < 3.0%.<br>Weak systems (SCR ≤ 20): voltage distortion limits by POI voltage (released as a chart).<br>Incremental limits: harmonics 2–4 and 6–50 ≤ 1.0% of nominal; 5th harmonic ≤ 0.3%.<br>Limits apply from 120 Hz to 3 kHz (50th harmonic).<br>Facility Owner shall disclose the electrical characteristics of any process causing cyclic active-power demand variability exceeding 5% of Maximum Rated Demand within any 10-minute period. |
| Ameren | IEEE 519 | Entity-specific voltage distortion limits:<br>69–161 kV→ Individual Harmonic Distortion: 1.5% , Total Harmonic Distortion: 2.5%<br>≥ 161 kV→ Individual Harmonic Distortion: 1.0% , Total Harmonic Distortion: 1.5% |
| Southern Co. | Southern Co. PQ Policy | May require individual harmonic current injections up to the 50th harmonic order at the POC. Specific numeric limits per Southern Company Power Quality Policy. |

*Voltage flicker*

Six entities address flicker, with IEEE 1453 as the common reference. The most common benchmarks are $P_{st}$ = 0.8 and $P_{lt}$ = 0.6. Southern Company and LIPA set stricter individual emission limits of Pst = 0.35 and $P_{lt}$ = 0.25 for service above 35 kV, while the total cumulative flicker that the system can tolerate from all sources combined at that location are 0.8/0.6.

**Table IX. Voltage Flicker Requirements Comparison**

| Entity | Reference Standard | $P_{st}$ Limit | $P_{lt}$ Limit |
|---|---|---|---|
| AEP | IEEE 1453-2022 | 0.8 | 0.6 |
| Dominion | IEC 61000-4-15 | 0.8 | 0.6 |
| Duke Energy | IEEE 1453-2015 | 0.8 | 0.6 |
| Ameren | IEEE 1453 | 0.8 | 0.6 |
| Southern Co. | Southern Co. PQ Policy | 0.35 | 0.25 |
| LIPA | IEEE 1453 | 0.35 | 0.25 |

*Voltage unbalance*

Voltage unbalance is the least commonly specified category. Only Duke Energy and Southern Company provide explicit numeric limits.

**Table X. Voltage Unbalance Requirements Comparison**

| Entity | User-Caused Limit | System Limit | Additional Notes |
|---|---|---|---|
| Duke Energy | ≤1% | ≤2% | Measured at POI for durations ≥1 min |
| Southern Co. | ≤1% | ≤3% | Measured at POI |
| LIPA | Balanced phases required | Not stated | Phase-to-phase or phase-to-ground voltage must not exceed 1.2 p.u. (instantaneous) or 1.1 p.u. for more than 1.0 s, measured over a 1/60-s window. |

## 2.7 Reactive power control

Table XI summarizes reactive power and power factor requirements. While LIPA requires a power factor of at least 0.98 and CAISO specifies a 0.97 lagging to 0.99 leading band, the 0.95 power factor threshold is the most common benchmark.

**Table XI. Reactive Power and Power Factor Requirements Comparison**

| Entity | PF Requirement | Reactive Compensation Notes |
|---|---|---|
| AEP | ≥ 0.95 lagging (near unity target) | Switched shunt capacitors are the common means<br>Requester responsible for voltage regulation on its side |
| Ameren | 0.95 lag to 0.95 lead | Controlled PF-correction devices required on customer side.<br>Cap/reactor switching voltage fluctuation limited to 3.3% normal, 5% single contingency. |
| CAISO | 0.97 lagging to 0.99 leading | Dynamic reactive-support requirements are under development. |
| Dominion | ≥ 0.95 lagging or 0.95 leading (near unity target) | Static capacitors, shunt reactors, and solid-state reactive compensation devices are identified tools. |
| Duke Energy | PF requirements by load level (table not yet published) | Duke can require PF correction equipment to compensate for reactive demands and to address reliability or operational issues identified in studies. |
| LIPA | ≥ 0.98 | Interconnecting entity expected to supply its own reactive power<br>During voltage disturbances (POI < 0.9 p.u.), reactive-current demand must not exceed 50% of Max Gross Demand current for >10 s or 100% for >1 s.<br>Sudden load discontinuation must not cause POI voltage to rise >5% for >1 s or >20% for >0.1 s. |
| SPP | NA | For IBR plants paired with Large Loads, if the plant cannot deliver both active and reactive power because of a current limit or apparent-power limit, and the applicable voltage is below 95%, preference shall be given to reactive power while minimizing reductions in active power. |

## 3. CONCLUSION

This paper presented a comparative analysis of large load interconnection requirements across eleven U.S. transmission entities. The analysis spanned seven core technical categories: large load definition, load data and modeling, fault ride-through, load curtailment and shedding, dispatchability, power quality, and reactive power control. The findings address many of the technical gaps in interconnection practice and reveal a landscape that is evolving rapidly.

The maturity assessment shows that LIPA now addresses all technical categories, at least partially, as of 2026. Dominion and Duke Energy stand out as the next tier, providing explicit or detailed requirements across nearly all categories. ERCOT, MISO, Southern Company, and SPP also demonstrate strong coverage. MISO's new Large Load Working Group framework fills many previous gaps, particularly in fault ride-through and post-fault recovery. Dispatchability, including ramp-rate constraints and load oscillation limits, is another emerging category, and Southern Company and LIPA are the only entities to publish a full quantitative framework for it. However, the two frameworks diverge substantially in their specific limits and measurement methods, so dispatchability remains a significant blind spot for grid stability.

One of the most important takeaways from this analysis is the growing realization that data centers do not have to be seen only as loads. Modern hyperscale data centers routinely deploy tens or hundreds of megawatts of diesel or gas-fired backup generators and, increasingly, utility-scale battery energy storage systems to ensure uninterruptible operation. If these resources are treated as a dispatchable grid energy source, the data center transforms from the grid's largest new burden into a distributed generation resource.